\pdfoutput=1
\documentclass{JINST}
\usepackage{upgreek}
\usepackage{url}
\usepackage{lineno}

\title{Further Developments in Gold-stud Bump Bonding}

\author{C.~Neher$^a$\thanks{Corresponding author.}, R.~L.~Lander$^a$, A.~Moskaleva$^a$, J.~Pasner$^a$, M.~Tripathi$^a$, M.~Woods$^a$\\
\llap{$^a$}University of California, Davis,\\

E-mail: \email{ceneher@ucdavis.edu}}

\abstract{As silicon detectors in high energy physics experiments require increasingly complex assembly procedures, the availability of a wide variety of interconnect technologies provides more options for overcoming obstacles in generic R\&D. Gold ball bonding has been a staple in the interconnect industry due to its ease of use and reliability.  However, due to some limitations in the standard technique, alternate methods of gold-stud bonding are being developed. This paper presents recent progress and challenges faced in the development of double gold-stud bonding and 0.5~mil wire gold-stud bonding at the UC Davis Facility for Interconnect Technology. Advantages and limitations of each technique are analyzed to provide insight into potential applications for each method.  Optimization of procedures and parameters is also presented.}


\begin{document}

\section{Introduction}

We have earlier reported~\cite{tripathi} on our development of gold-stud bump bonding, for applications in high energy physics experiments, at the UC Davis Facility for Interconnect Technology~\cite{ucdfit}. This technique eliminates the need for photolithography on a single die (individual prototype IC chips), which is typically required in order to provide suitable under-bump metallization.  The gold-studs break through the oxide layer on the die pads, most commonly made of aluminum, and make good ohmic connections. The studs may be mated to opposing aluminum pads using thermo-sonic bonding or thermo-compression. We utilize the latter. However, thermo-compression still requires that the pads on the mating part have a top layer of gold in order to achieve a good attachment. The preparation of this gold surface, either via electro-less processes or sputtering, still requires photolithography steps. In this paper we report on a ``double gold-stud'' process, which entirely removes any photolithography steps from the bump bonding process. Another development that we have pursued is aimed at reducing the gold-stud size, which would enable us to employ this technique on parts with a smaller pad pitch.  We present studies of gold-stud compression and a demonstration of bump bonding performed using smaller diameter gold wire.

\section{Gold-stud Bump Bonding}
\begin{figure}[tbp]
\begin{center}
\includegraphics[width=.45\textwidth]{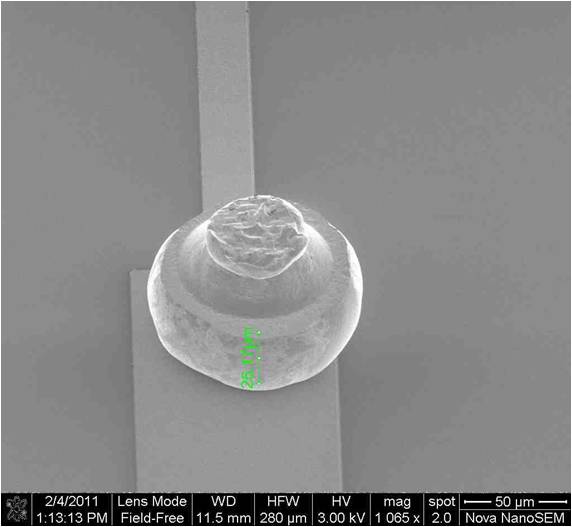}
\includegraphics[width=.45\textwidth]{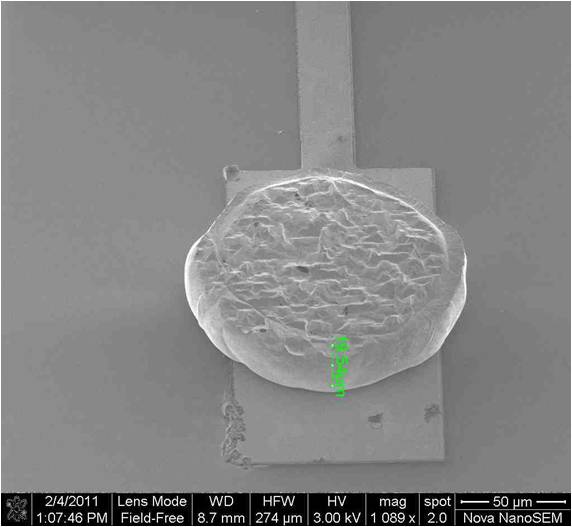}
\caption{An SEM image of a gold stud bump before (left) and after (right) coining. Coining is performed by compressing the studs under a flat piece of silicon using 140~g/stud. The result is a flattened bump, allowing for easier alignment with bumps on the mating chip.}
\label{fig:sem_coining}
\end{center}
\end{figure}

Gold-studs are formed using a Westbond ball bonder~\cite{westbond}, which is equipped with a ceramic capillary. Under normal operations, a 1~mil (25.4~$\upmu$m) gold wire is threaded through the capillary. A high voltage arc, applied to the tip of the wire by a pneumatic arm, melts the gold and forms a small ball at the tip. The capillary, now carrying the gold ball at its tip, is brought to the surface of the device being studded, and after reaching a threshold of pressure, uses ultrasonic energy to create friction between the ball and the substrate to form a metallic bond.  The capillary is then retracted, breaking off the wire and leaving the stud on the pad, as shown in Figure~\ref{fig:sem_coining}~(left). Once the studs are in place, they can be flip-chip bonded to mating gold pads on a second chip. We accomplish this on our ``Finetech Pico ma'' aligner-bonder~\cite{finetech}, using thermo-compression with a force of 160~g/stud at a temperature of 320~$^\circ$C, which provides optimal results in yield and shear strength~\cite{tripathi}. Alternately, the gold-stud can be first flattened through a process called ``coining'', a technique that involves compressing the studs over a flat piece of silicon with roughly 140~g/stud, at room temperature.  A typical coined stud is shown in Figure~\ref{fig:sem_coining}~(right).

\begin{figure}[tbp]
\begin{center}
\includegraphics[width=.9\textwidth]{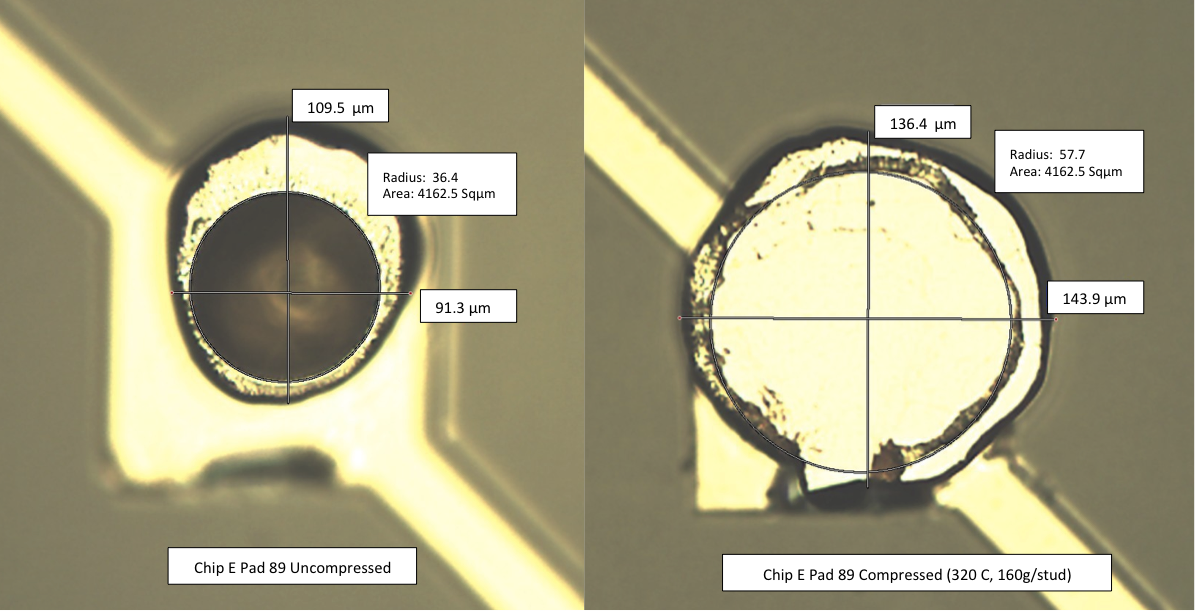}
\caption{A comparison of a gold-stud before (left) and after (right) thermo-compression. The left image shows the formation of a ``head'' and ``collar'' with different dimensions. The right image shows the merging of the head and collar.}
\label{fig:collar}
\end{center}
\end{figure}

The uncompressed gold-studs contain two features: a highly regular circular ``head'' atop an irregular semi-elliptical ``collar''. The head is consistently 72$\pm$1~$\upmu$m in diameter, reflecting the capillary bore, while the collar is 104$\pm$6~$\upmu$m in diameter. During bonding (160~g/stud, 320~$^\circ$C) the head is compressed into the collar to form a plateau 143$\pm$12~$\upmu$m in diameter. Comparison for a typical case is shown in Figure~\ref{fig:collar}. The error on the alignment of the stud and pad centers is 16$\pm$7~$\upmu$m. Combining the errors, we determine the stay-clear for one gold bond to be 164$\pm$21~$\upmu$m, which sets the minimum pitch size for this technique (using 1~mil wire).

Gold-stud bonding is a well understood interconnect technology and can be used to assemble a wide range of prototypes with reliably low resistance interconnects, making it an invaluable tool for detector development and integrated circuit research. Because ball bonding does not require photolithography or a specific metal stack, a trained user can prepare a sample for bonding in a relatively short amount of time.  The typical resistance of a gold-stud bond is less than 1 m$\Omega$.  However, we consider a bond to be acceptable for our applications if it has a resistance < 0.5 $\Omega$.

There are a few drawbacks to using gold-studs as an interconnect technique. The typical pressure and temperature required for thermo-compression can be hazardous to sensitive circuitry. Also, devices that require a very high numbers of gold-studs will consequently require a very high pressure to bond, which can exceed the capability of the aligner bonder. Similarly, the high temperature required can weaken bonds of other interconnects, such as solder bumps, if they were in place before the gold bond. Lastly, gold balls formed from the usual 1~mil wire expand out to around 140~$\upmu$m during thermo-compression, implying that shorting is a potential problem for devices with small gaps between pads.

The UC Davis Facility for Interconnect Technologies (UCD-FIT) has been developing methods for removing the negative aspects of gold-studs, in order to make them a viable option for a wide variety of prototyping needs. This paper explores double gold-stud bonding as a method of reducing the bonding stresses, and 0.5~mil gold wire studs as a technique for bonding smaller devices.

\section{Development of Double Gold-stud Bonding}
\begin{figure}[tbp]
\begin{center}
\includegraphics[width=.9\textwidth]{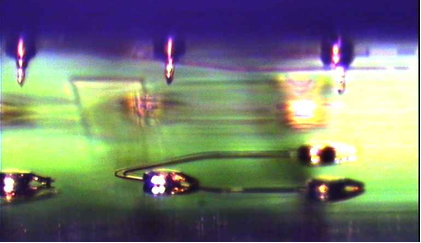}
\caption{An example of double gold-stud bonding. The top chip has gold studs that have pointed ends, while the bottom chip has studs that have been coined into flat tablet shapes.  This process requires lower pressure and temperature than single gold stud bonding. Also, it works directly with aluminum pads with no need for under-bump metallization.}
\label{fig:during_bond}
\end{center}
\end{figure}

Double gold-stud bonding has been developed in much in the same way as single gold-stud bonding, with the primary difference being that gold-studs are placed on both the die and the substrate rather than on just one device.  Stud-to-stud thermo-compression completes the bond. If the contact during the bonding were to be made between two semi spherical gold-studs, there would likely be slippage or inadequate contact due to minor misalignment. To work around this, the studs on one or both devices can be coined. This creates a better surface for contact, which greatly improves bond yield. Once this step is completed, the devices to be bonded are aligned and bonded through thermo-compression.  Figure~\ref{fig:during_bond} shows a pair of chips just prior to attachment --- in this instance the studs on the bottom chip are coined, while the upper studs have visible tails from the gold wire.

In order to test the viability of double studding as an interconnect technique, the parameter space of temperatures and pressures for bonding was explored. To test the resistance of the double stud bonds, dummy chips with an array of 20 pads, each with traces allowing a four-point (Kelvin) measurement of the bond resistance were used. Because gold studs are formed with a manual gold ball bonder, twenty pads provide sufficient statistics for our investigations without demanding a superfluous amount of time on the machine. Both chips had gold-studs placed on the bond pads, and both sides were coined at 140~g/stud. Once the samples were properly prepared, we bonded pairs of chips under varying conditions using the Finetech flip-chip bonder. 

Starting with our preferred bonding parameters for single gold-stud bonding (320~$^\circ$C and 160~g/stud), the first set of bonds was carried out with a constant temperature of 320~$^\circ$C and at various pressures decreasing at 10~g/stud intervals from 160~g/stud until the bond yield was no longer 100\%. After these tests were completed, the next set of bonds was carried out under a constant pressure of 160~g/stud, and with each successive bond 10~$^\circ$C lower in temperature until the bond yield dropped below 100\%. The final set of bonds was carried out with both the temperature and pressure being lowered in increments of 20~$^\circ$C and 10~g/stud, respectively. Once the bonds were completed, those with 100\% yield were tested for shear strength using a force meter pulling the chips apart until the bond was broken.  We define 100\% yield as an assembly in which none of the bonds has a resistance > 0.5 $\Omega$.

\begin{figure}[tbp]
\begin{center}
\includegraphics[width=.95\textwidth]{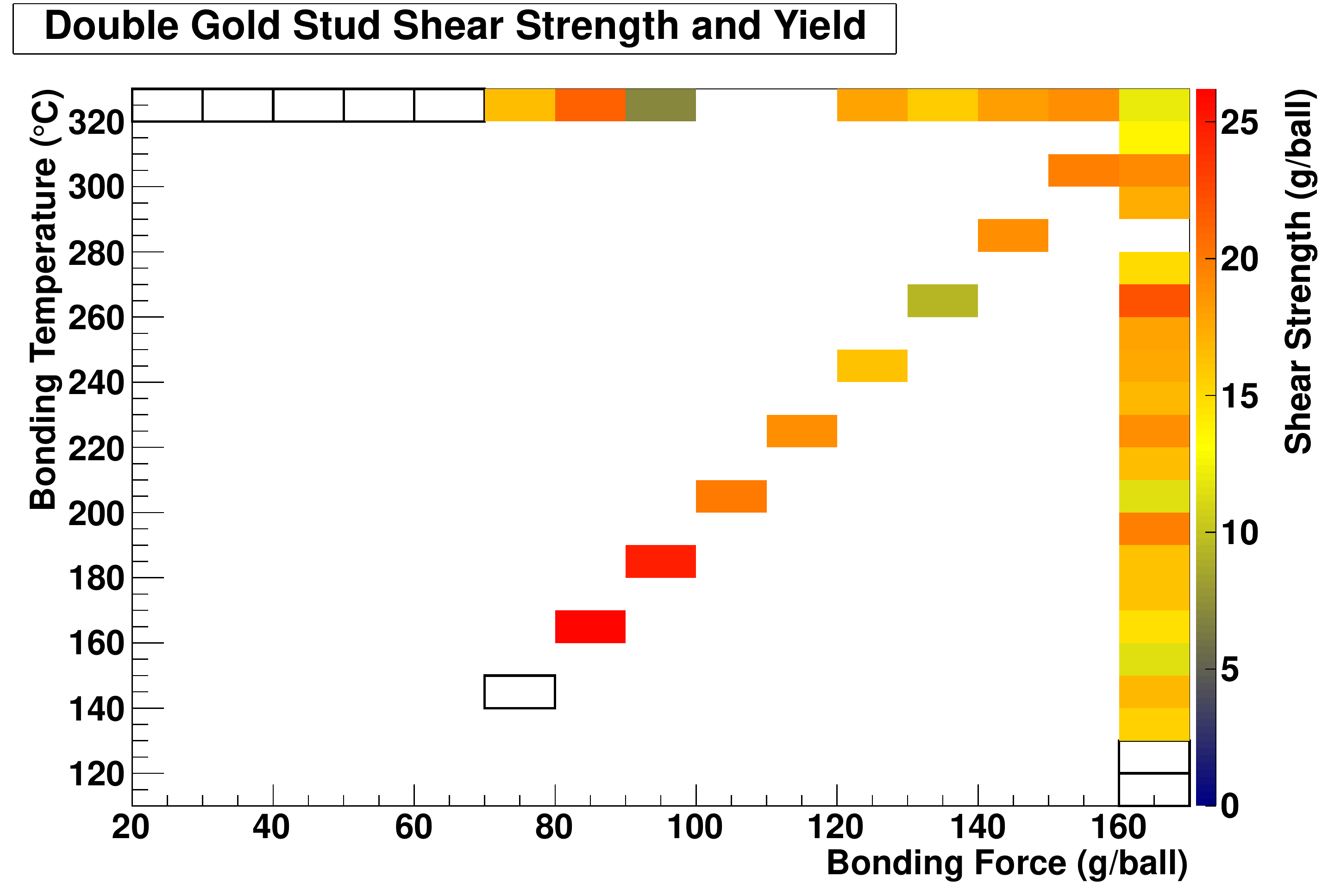}
\caption{A graphical representation of the results from the Bonding Parameter Exploration tests of double gold studs. The plot shows the amount of force and heat used during each bond, with the colored squares corresponding to bonds with a yield of 100\%, and the clear squares with black outlines being bonds with less than 100\% yield. The color of the squares corresponds to the amount of force required to shear the bond formed under those parameters. The blank spots at 100-110~g/stud at 320~$^\circ$C are bonds that fell apart before the shear test.}
\label{fig:yield_and_shear}
\end{center}
\end{figure}

The results from this exercise are displayed in Figure~\ref{fig:yield_and_shear}. Boxes filled in with color represent trials with 100\% yield. For the tests with constant temperature of 320~$^\circ$C and variable pressures, it was found that bonds made with greater than 70~g/stud had 100\% yield. When the pressure was kept constant at 160~g/stud and the temperature was varied, it was shown that bonds above 130~$^\circ$C were 100\% successful. The tests with a combination of variable pressures and temperatures showed that bonds at 160~$^\circ$C and 80~g/stud had 100\% yield.

\begin{figure}[tbp]
\begin{center}
\includegraphics[width=.445\textwidth]{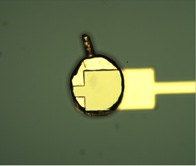}
\includegraphics[width=.5\textwidth]{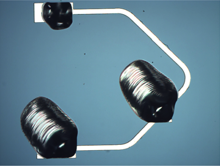}
\caption{Images showing 2 dummy chips after being sheared apart. The left image shows a gold stud that has pulled of a piece of the pad and trace from its complementary chip. The right image shows a dummy chip that has had chunks of silicon torn off out of the chip at the bonding locations.}
\label{fig:post_shear}
\end{center}
\end{figure}

The pulling tests revealed that most bonds had shear strengths of roughly 15~g/stud, though the bonds with a combination of low temperature and pressure had shear strengths closer to 25~g/stud. This unexpected behavior may be explained by another factor. In most cases the values are lower limits on the stud-to-stud interface's actual shear strength, as most breaks occurred at the stud to chip connection rather than the stud-to-stud contact. This is illustrated in Figure~\ref{fig:post_shear}. In many cases during the shear test the studs pulled off parts of the underlying pads and even chunks of silicon from the chip, indicating that double stud connections are quite strong.  Hence, we conjecture that as the temperature of the bonding process increased, the strength of the pad to oxide substrate connection was somehow weakened.

\begin{figure}[htbp]
\begin{center}
\includegraphics[width=.8\textwidth]{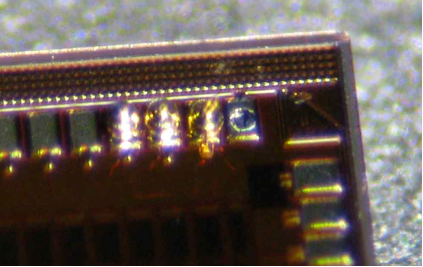}
\caption{Image showing gold studs from a 1~mil wire placed on the corner of a PSEC chip. The size of the studs overhangs the edges of the pads, which puts them at risk of shorting.}
\label{fig:big_psec}
\end{center}
\end{figure}

\section{R\&D Towards Smaller Gold-studs}
Development for this project began during our work with the PSEC waveform sampling ASIC for the Large Area Picosecond Photo-Detector (LAPPD) collaboration~\cite{abrams,wetstein}. The periphery of the PSEC chip contains 118 square pads, each 62~$\upmu$m on a side, with a pitch of 130~$\upmu$m.  As shown in Figure~\ref{fig:big_psec}, gold-studs formed using 1~mil wire are too large and would cause shorts during thermo-compression. In response to this problem we began using 0.5~mil (12.7~$\upmu$m) gold wire to form smaller gold studs. 

\begin{figure}[tbp]
\begin{center}
\includegraphics[width=.9\textwidth]{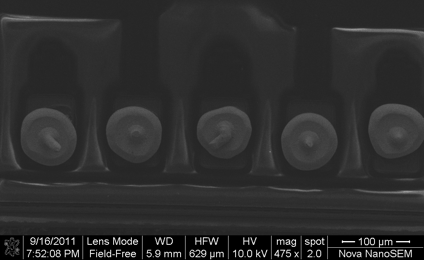}
\caption{SEM image showing a close-up on 0.5~mil wire studs placed on the pads of a PSEC chip. The smaller diameter keeps the edges of the gold-studs within the pad boundaries, eliminating the risk of shorting.}
\label{fig:small_psec}
\end{center}
\end{figure}

The procedure with 0.5~mil wire uses the same Westbond Ball bonder as the 1~mil wire, albeit with a smaller capillary. The thinner wire also necessitates different settings for the ultrasonic power and studding force. We have determined that using an ultrasonic energy of 400 for 45~ms yields a greater fraction of properly formed studs, and with greater adhesion strength to the substrate. Using these settings, we have been able to achieve stud collar sizes of 87$\pm$4~$\upmu$m, significantly smaller than 104$\pm$6~$\upmu$m from the 1~mil wire. Figure~\ref{fig:small_psec} has an example of such smaller gold-studs, showing a scanning electron microscope (SEM) picture of the gold studs. We have also performed a successful thermo-compression bond of a prototype PSEC using the pressure and temperature typically used for 1~mil wire studs.  This assembly is being tested by the LAPPD collaboration.

Despite achieving smaller gold-studs, we are aware of difficulties with the 0.5~mil wire that must be addressed. In particular, the gold wire is much more difficult to handle compared to its thicker counterpart. The wire is prone to bending and kinking while threading through the capillary, which results in far more time being spent preparing the bonder. Faults in the machine are also more likely to occur. The specific sonication properties need to be more thoroughly quantified because higher studding forces flatten the gold studs into larger diameters. Therefore, further work is required to find the lowest amount of force that still forms a strong stud-to-pad connection. 

\section{Conclusions}
We have developed a double gold-stud process, which should find application in situations where preparation of pad surfaces is not easily accomplished. It provides attachments with 100\% yield over a wide range of temperature-pressure parameter space.  The lowest observed values were determined to be 160~$^\circ$C and 80~g/stud.  The strength of stud-to-stud attachment is at least 15~g/stud, and most likely as high as 25~g/stud. We have made progress in the use of 0.5~mil gold wire in order to form studs with a smaller diameter, to be used in applications with a finer pitch.

\acknowledgments

This work is supported by a grant from the US Department of Energy, Office of High Energy Physics, under section ADR KA150301.

\end{document}